\documentstyle[aps,prl,multicol,epsfig]{revtex} 
\begin{document}

\title{Effect of coulomb long range interactions on the Mott
Transition}
\author{ R. Chitra and G. Kotliar}
\address
{Serin Physics Laboratory, Rutgers University,
Piscataway, NJ 08855, USA}
\date{\today}
\maketitle
\begin{abstract}

We reconsider the  Mott transition problem
in the presence of long range Coulomb interactions.
Using an extended DMFT, that  sums  an  important
class of diagrams absent in ordinary DMFT, we show
that in the presence of Coulomb the Mott transition
in two and three dimensions is  discontinuous 
as envisioned  by Mott.


\end{abstract}

\begin{multicols}{2}


The study of Metal Insulator transitions\cite{nrev} (MIT) seen in a host of interacting
 electronic systems,
has been one of the most challenging problems in solid state physics.
These transitions can  result from simple band
structure or from electronic correlations. 
The former  occurs even in nearly non-interacting systems if the
conduction and valence bands split as some parameter like composition
is varied. 
The correlation driven transition i.e., Mott transition, is however, the most interesting.
The nature of these transitions depend on the interplay of
band structure, magnetism and
electronic correlations.
 The simplest scenario for a MIT was first
put forth by Mott \cite{mott}.
Mott suggested that a crystalline array of hydrogenic atoms i.e., atoms with just
one electron in their outermost shell,
shows
a zero temperature transition from a metal to an insulator as the
density is decreased or effectively, as the distance between atoms increases \cite{hub}.
Later, other possibilities for the MIT were put forth, namely, Slater's
band transition \cite{slater} where the system becomes insulating because of the
doubling of the unit cell due to antiferromagnetism and  the Brinkman-Rice
scenario within the Hubbard model \cite{br}  which  is characterised
by a  strong mass enhancement.

Mott's original argument depended on the presence of coulomb long range
interactions and pointed out that 
since the number of "free" electrons could only
vary discontinuously across the transition, the transition is
necessarily first order.
We recapitulate Mott's ideas below:
the main premise was that, due to the long range nature of the interactions,
the electrons and the holes in the two Hubbard bands will always form
bound states and not exist independently.
The  condition for the formation of atleast one bound state, however, 
depends on the screening of the
coulomb interaction.  Using Thomas-Fermi
 estimates for screening by $N$ electrons per
unit volume, Mott showed that no bound states exist provided
\begin{equation}
N^{1/3} a_H > 0.4 
\label{crit}
\end{equation}
where $a_H$ is the Bohr radius. This implies that
the system is metallic as long as the density satisfies (\ref{crit}).
For densities where (\ref{crit}) is violated, the electrons and holes in
the Hubbard bands always form bound pairs resulting in an insulating
behavior. This implies that the number of carriers jumps at the
transition. Since this results in a kink in the free energy, the
transition is discontinuous and first order.  
For an illuminating discussion of these ideas see Ref.\onlinecite{giamarchi}.

In contrast, various studies of the Hubbard model 
\cite{hub,br,rev} indicate that the transition is continuous.
The Hubbard model for interacting electrons has long been used as
a prototype to describe various aspects of real systems\cite{nrev}.
This model  contains a term which
describes the hopping of electrons between different atomic sites
and another describing 
the coulomb repulsion felt by two electrons on the same atom.
 In the past few years, the nature of
the MIT in this model has been clearly  elucidated  using the
Dynamical Mean Field theory (DMFT) in the limit of infinite lattice coordination $d$
\cite{rev}.
It was shown that the MIT was characterised by a continuous vanishing
of a Kondo like resonance at the Fermi level in the metal 
 at the transition, leading to
an insulator with a preformed gap. This led to the effective mass of the
quasi-particles 
and hence the linear coefficient
diverging from the metallic side at the transition \cite{br,rev}.
Despite the discontinuous opening of  the  Mott Hubbard gap
the destruction of the metal  
at zero temperature treated with DMFT,  is a   continuous  one.

The purpose of this work is
to incorporate some of the effects
of the long range coulomb interactions
and Mott's ideas into the  framework of Dynamical Mean Field Theory.
For this purpose, we explore
the effects of the long  range interaction on the MIT seen in the
single band Hubbard model \cite{hub}, using a simple extension
of DMFT \cite{kajueter,Si}.  It was shown, in a model of spinless fermions,
that this approach captures important $ 1 \over d$ corrections
\cite{kajueter}.  This method was independently developed in  Ref.\onlinecite{Si}
and it was applied to the problem of the  breakdown of Fermi liquid theory.
We first describe  the approach 
by isolating  a class of diagrams
which can be formally controlled by scaling the interactions 
 and the kinetic energy  appropriately
and which can be summed using impurity models. We then demonstrate that
when this extended DMFT is applied directly
to a $3$ or $2$ dimensional  lattice with  interactions having
the  coulomb form, 
it changes  the Mott transition which was continuous in ordinary DMFT,
to a discontinuous  
first order transition as envisioned by Mott. 

The  effective hamiltonian  we  use to describe
our system is  a generalization of the Hubbard hamiltonian 
\begin{eqnarray}
H=&&  
\sum_{\langle ij\rangle \sigma} t_{ij} c^+_{i\sigma} c_{j\sigma} + h.c.  
+ \sum_{i\sigma} \mu c^+_{i\sigma}  c_{i\sigma}   \nonumber  \\
&&+
\sum_i U n_{i\uparrow}
 n_{i\downarrow} +\sum_{i\ne j,\sigma \sigma^{\prime}} V_{ij}:n_{i\sigma}n_{j\sigma^{
\prime}}:
\label{ham}
\end{eqnarray}
\noindent
The first term is the hopping matrix element for  an electron from site $i$
(representing the ion at $R_i$)
to its  neighboring site $j$, $\mu$ is the chemical potential,  $U$  is the coulomb repulsion
felt by the electrons   when they are on the same atom ($i=j$) and the
normal ordered last term is the   coulomb interaction 
between  electrons  on different atoms.  
The coupling constants are given by  overlap
integrals involving   a set of chosen basis vectors like the
 Wannier or Hartree
basis.
For example, the on-site interaction $U$ is given by
\begin{eqnarray}
U& = & \int d{ r} d{ r}^\prime \vert u(r)\vert^2
\vert u(r^\prime)\vert ^2 {{e^2} \over {\vert r -r^\prime \vert}} \\ \nonumber
V_{ij}& =&  \int dr dr^{\prime} 
\vert u(r-R_i)\vert^2
\vert u(r^\prime-R_j)\vert ^2 {{e^2} \over {\vert r -r^\prime \vert}}
\end{eqnarray}
\noindent

The hamiltonian (\ref{ham}) is now studied in the dynamical mean field approximation.
The first step in the DMFT is to scale the parameters in the 
 large $d$ limit appropriately such that the corresponding energy terms
remain finite. In addition, the scaling should be chosen such that the  
terms of interest remain relevant in the large $d$ limit.
We adopt the following scaling:
$t_{ij}$ is scaled as 
${\sqrt{d}}^{-\vert i-j\vert}$,
$U \to U$
and  
$V_{ij}$ is scaled by ${\sqrt{d}}^{-\vert i-j\vert}$.
The leading diagrams can then be summed up 
using  the cavity method described in
Ref.\onlinecite{rev}. 
This scaling is a prescription for choosing  a set of diagrams that contribute
to the self energy and hence the Green's function. 
DMFT  prescribes a scheme for choosing 
different sets of diagrams that contribute to the different Green's functions.
In the case of the Hubbard model ($V=0$), this corresponds to retaining
only skeleton diagrams constructed from $U$ and the local Green's functions
$G_{ii}$ in the self energy, resulting in the self energy $\Sigma_{ij}$ being local.
\begin{figure}
\centerline{\epsfig{file=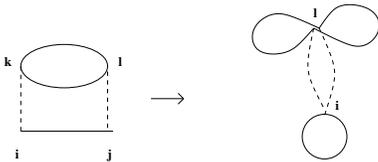,angle=360, width=5cm}}
\narrowtext
\caption{Reduction of the self-energy diagrams in the limit of
infinite $d$.}
\label{fig1}
\end{figure}
\noindent
In the presence of the longer range interaction, the above scaling retains
all skeleton diagrams constructed using the local Greens function
$G_{ii}$ and the interaction vertices $U$ and $V_{ij}$,
such that  every
point $i$ which has a vertex $V_{ij}$ originating from it
 has another vertex $V_{ki}$
terminating at it. 
In effect,  this corresponds to replacing the $U$ by an $U_{eff}$
 in the local self energy
evaluated for a Hubbard model.
An example of the diagrams retained is shown in Fig.\ref{fig1}.
We mention that  the Hartree term (arising only from $U$)  is generated
 in the cavity method. 
 Depending on the  basis  chosen to derive the effective
model parameters (\ref{ham}) care should
be exerted to see that the Hartree term is not double counted. 
We neglect the Fock term in the ensuing calculations 
, since
 the Fock term is of higher order
in  $ 1 \over d$  
 than the Hartree term. 

Using the cavity method and integrating out all sites save a chosen site or cavity $o$, we obtain the following
local effective impurity action with retarded interactions 
\begin{eqnarray}
S_{eff}= &&\int d\tau d\tau^{\prime} \sum_{\sigma}c_{0\sigma}^+ (\tau) {\cal G}_0^{-1}(\tau -\tau^{\prime})
c_{0\sigma}(\tau^{\prime})  \nonumber \\
&&+ U n_{0\uparrow}(\tau)  n_{0\downarrow}(\tau) -
 \sum_{\sigma \sigma^\prime}  n_{0\sigma}(\tau) 
 \Pi^{-1}_{0} (\tau -\tau^{\prime}) 
n_{0\sigma^{\prime}}(\tau^{\prime})  
\label{himp}
\end{eqnarray}
\noindent
where
\begin{equation}
{\cal G}_0^{-1}= \partial_\tau - \sum_{ij}t_{0i}t_{0j} G_{ij}^{(0)}
(\tau -\tau^{\prime}) 
\label{g0}
\end{equation}
\noindent
and the retarded interaction
\begin{eqnarray}
\Pi_{0}^{-1} (\tau -\tau^{\prime})&=&  
 \sum_{ij}V_{0i} V_{0j} \Pi_{ij}^{(0)}(\tau-
\tau^{\prime})  \nonumber \\
&= & \sum_q 
 V(q)^2 \Pi^{(0)}(q,\tau-\tau^\prime)
\label{screen}
\end{eqnarray}
\noindent
with $G_{ij}^{(0)}$ denoting the single particle Green's function and 
$\Pi_{ij}^{(0)}= \langle\sum_{\sigma\sigma^\prime} n_{i\sigma}(\tau) n_{j\sigma^\prime
}(\tau^{\prime})
\rangle_0$. The superscript $(0)$ implies that
the quantities have been evaluated in the system from which the site
$0$ and all the links to this site  have been removed.
These in turn are 
 related to quantities evaluated on the entire lattice. 
Summing over the sites leads to ${\cal G}_0$ and $\Pi_0$ 
being determined by self-consistent equations involving
local quantities evaluated on the full lattice.
Note that the effect of all non-local quartic interactions 
is to  dynamically screen the on-site repulsion, 
with the screening  potential given by (\ref{screen}).  
These equations depend on the
nature of the lattice used. 
For  example, on the Bethe lattice, 
$G_{ij}^{(0)}= G_{ij}$ and   $\Pi_{ij}^{(0)}= \Pi_{ij}$.
Retaining only the on site and nearest neighbor
interactions, the self-consistency condition
that should be satisfied by the retarded
interaction on the Bethe lattice 
takes the  simple form
$\Pi_{0}^{-1}=  V^2 \Pi_{loc}$, 
where $\Pi_{loc}$ is the local density density correlator.
For arbitrary lattices  and interactions
(i.e. general  $t_{ij}$ and $V_{ij}$) 
the  (extended) dynamical mean field
equations can be derived by generalizing the discussion
presented here, to the case where  the hoppings  $t_{ij}$ are 
scaled by ${\sqrt d}^{- \vert i-j \vert}$.
 
To obtain the Green's functions of (\ref{himp}), 
we  now define  certain irreducible quantities.
The lattice Green's functions, can be expressed in 
terms of a self energy $\Sigma$ which is two particle irreducible and
 which  becomes local in the limit of infinite dimensions.
\begin{equation}
G(i \omega_n , q)=  {1 \over {i\omega_n -\epsilon_q - \Sigma( i\omega_n )}}
\end{equation} 
\noindent
where $\epsilon_q$ is the dispersion on the lattice and $i\omega_n$ are
the Matsubara frequencies.
Similarly, the density 
density correlator $\Pi (q,i\omega_n)$ on the lattice
defines an irreducible part ${\tilde \Pi}$ via the Dyson 
equation
\begin{equation}
\Pi (q,i\omega_n) = {{\tilde \Pi} \over { 1+ V(q) {\tilde \Pi}}}
\label{plat}
\end{equation}
\noindent
where $\tilde \Pi$ is the sum over  all
polarization  diagrams  constructed with the full $G$ and 
interaction vertices $V(q)$  such that all the diagrams
are irreducible with respect to $V(q)$.
Within the infinite $d$ approximation, since all vertex functions
become independent of momenta, $\tilde \Pi$ also becomes independent of
momenta.
The impurity model
(\ref{himp})  allows us to compute all local quantities,
 in particular
  $\Sigma $ and  ${\tilde \Pi}$ and hence,  
 the  local density 
density correlator   $\Pi_{loc}$ and the local
one particle 
Green's function $G_{loc} $ 
as functionals of the Weiss fields  ${\cal G}_0$ and 
$\Pi_0$.
\begin{equation}
\Pi^{-1}_{loc}\{\Pi_0,{\cal G}_0\}(i\omega_n)= 
{\tilde \Pi}^{-1}\{\Pi_0,{\cal G}_0\}(i\omega_n) - 
\Pi_0^{-1}(i\omega_n)  
\label{pimp}
\end{equation}
\noindent
and
\begin{equation}
\Sigma\{\Pi_0,{\cal G}_0\}(i\omega_n)= {\cal G}_0^{-1}(i\omega_n) - 
G_{loc}\{\Pi_0,{\cal G}_0\}(i\omega_n) 
\end{equation}
\noindent
Using (\ref{plat}) and (\ref{pimp}), we can eliminate
$\tilde \Pi$ to obtain a
self-consistent equation for $\Pi_{loc}=  \sum_q \Pi(q,i\omega_n)$
\begin{equation} 
\Pi_{loc}\{\Pi_0,{\cal G}_0\}= \sum_q {1 \over { \Pi_{loc}^{-1}\{\Pi_0,{\cal G}_0\} -\Pi_0^{-1} +V(q)}}
\label{sceq}
\end{equation}
\noindent
Similarly, we obtain an equation  for $G_{loc}$
\begin{equation}
G_{loc}\{\Pi_0,{\cal G}_0\} =  \sum_q {1 \over
 { i\omega_n -\epsilon_q - \Sigma\{\Pi_0,{\cal G}_0\}}}
\end{equation}
\noindent

These equations where derived by an appropriate scaling of the interactions
and the hopping elements in the original model so as to obtain a well defined
limit of large coordination.
In the spirit of DMFT \cite{rev},
we can, however, regard  these equations as defined  on a finite dimensional lattice by
replacing $V$ by the usual coulomb interaction on the lattice. 
We use these equations  to make qualitative predictions of the effect
of the coulomb long range interaction  in finite dimensions, 
on the order of the
Mott transition seen.

In earlier  DMFT studies of the  Hubbard model
the  continuous Mott transition was
found to occur at a critical $U=U_{c2}$
\cite{marcelo} and 
was signaled by the vanishing of a Kondo like resonance 
 at $\omega=0$ at $U_{c2}$.
The continuous character of that transition was established using
the projective self consistent method \cite{goetz}.

We briefly review the projective method below.
Then, using  the results of Ref.\onlinecite{goetz},
we study  how 
 long range Coulomb interactions 
treated
within the extended DMFT,
modify the previous discussion.
The projective method 
uses the separation of two energy scales that exists in the metallic
phase  close to the MIT i.e., $wD$ where $w$ is
 the weight of the Kondo resonance seen in the metal 
and  scale of the Hubbard bands $U$. $D$ is the half band-width. 
 The  high energy scales are then eliminated to obtain an effective
theory governed by one low energy scale which is
 $w$,  which goes to zero at
the MIT\cite{rev}.  Using this effective low energy model, which is a Kondo
model of an impurity spin interacting with a bath of electrons,
 we can obtain the free energy or the ground state energy 
of the lattice problem close to the transition point. This free energy
correctly describes the low energy and coherent part of the spectra.
To order $w$, the low energy Kondo problem  has been derived in
Ref.\onlinecite{goetz} and is given by
\begin{equation}
{\cal H} = w\Gamma {\bf S}.{\bf s}_L + {\cal H}_b 
\label{kondo}
\end{equation}
\noindent
Here  ${\cal H}_b$ describes a band of  low energy conduction electrons  
 and ${\bf s}_L$ represents the local spin operator of 
these electrons. ${\bf S}$ is the impurity spin and the Kondo coupling
$\Gamma$ is determined by matrix elements in 
 the high energy sector (comprising
the two Hubbard bands). 
The self-consistency conditions translate to conditions on the expectation
value  $\langle {\bf S}.{\bf s}_L \rangle $. 
The high energy sector is an insulator and has a spin doublet
 ground state  which is
separated from the excited states by a large  gap. 
Using the results of Ref.\onlinecite{goetz}, the simplest 
approximation to the high energy sector of the Hubbard model yields
$\Gamma = \frac{2D^2}{U}$.
Since we are interested only in the qualitative features of the transition,
we replace the bath of electrons by a single electron.
Therefore, using this toy model,
the energy of the Kondo model is 
\begin{equation}
E_K= - \frac{3}{4}w \Gamma
\label{energy}
\end{equation}
\noindent
 Using the results of \cite{note},
and taking into account the kinetic energy of the lattice which is positive
and of order $w$,
the ground state energy of
the lattice model has the following expansion in terms of $w$
\begin{equation}
E_0 = (\alpha - \beta \frac{2D^2}{U}) w + \gamma w^2 +  {\rm
corrections}
\label{E0}
\end{equation}
\noindent
Here $\alpha,\beta,\gamma$  are all positive.
The transition is determined by the $U$ at which the coefficient of the
term linear in $w$ vanishes.  Since the energy minimum still occurs at
$w=0$, this transition is continuous. 

Though the DMFT and the self-consistent projective method are both formulated
in infinite dimensions, it is nonetheless known that they  the
capture some aspects of the 
physics of systems in finite dimensions rather well. 
We therefore, use the projective method to make some predictions about the
transition in three dimensions.
First, in the presence of the long range coulomb interaction,
there are corrections to (\ref{E0}), because of the screening of
the on-site coulomb interaction given by (\ref{screen}). 
This implies that $U$ in (\ref{E0}) has to replaced by an $U_{eff}$.
For the free energy (\ref{E0}), it is sufficient to
consider  the  effective static on-site  repulsion.
Taking  $\Pi^{(0)}= \Pi + \rm{ sub-leading corrections}$ in (\ref{screen})
and using the fact that 
at zero frequency, $\tilde \Pi$  in (\ref{plat}) is  by definition
 the compressibility $\kappa$ of
the system, 
one obtains from (\ref{screen})
\begin{equation}
U_{eff} = U - \sum_q V^2(q)  \frac{\kappa}{ 1+\kappa
V(q)} 
\label{ueff}
\end{equation}
\noindent
Note that $V(q)$ is the Fourier transform of the coulomb potential in $d$ dimensions. 
Since the   effective interaction remains
finite at the transition, the leading critical behavior of the 
compressibility  $\kappa$ is the same
as that found in the Hubbard model
and $\kappa$  is given by  its value in the Hubbard  model \cite{goetz,rev}
\begin{equation}
\kappa \propto \frac w{D} 
\label{kappa}
\end{equation}
\noindent
Note that $\kappa$ is finite in the metal 
and goes to zero smoothly at the transition. 
This implies that the effective repulsion seen by the electrons in the
metallic phase is reduced. In the insulator
$\kappa=0$ and  the bare on-site repulsion is not
screened at all. This is physically correct because one does expect the
itinerant electrons in the metallic phase to better  screen  the 
coulomb repulsion.
 Using (\ref{kappa}) in (\ref{ueff}), and performing the $q$ sum in $d=3$
we find that the effective repulsion felt on-site is 
\begin{equation}
U_{eff}= U- a w^{1/2} 
\label{ueff3}
\end{equation}
\noindent
where $a$ is some positive constant.
The screening is proportional to $\log w$ in two dimensions. 
Substituting (\ref{ueff3}) in (\ref{E0}), we see that the
 expansion  of the energy  in terms of the low energy weight $w$ has
non-analytic terms  i.e.,
\begin{equation}
E_0 = (\alpha - \beta \frac{2D^2}{U}) w - a\beta \frac{2D^2}{U^2}
w^{\frac3{2}} + \gamma w^2 + \rm {correction}
\label{Ena}
\end{equation}
\noindent

Notice that the parameter $ w$ in the projective self consistent method,
neatly embodies the notion of "number of free carriers" in Mott's
original number.
The  non-analyticity in (\ref{Ena}) arises purely from the self-consistency
conditions (\ref{sceq}) when 
 coulomb  interactions are taken into account. 
We now see that at the transition, though the linear term still vanishes,
the minimum of the energy  which formerly was at $w=0$ is  now
shifted to a non-zero
value because of this non-analytic term.  
  This is a feature of 
a first order transition! 
Hence, we see that taking into account the
long-range nature of the coulomb interaction does make the Mott transition
first order in $d=3$. 
Regarding the effects of coulomb interactions in two dimensions,
we find that
the  screening results in a term proportional to $w \log w$ in the
free energy (\ref{E0}),
again making the Mott transition discontinuous.
Even-though the DMFT is too crude an approximation in $d=2$ where spatial
fluctuations are large, 
 we argue that it still captures some of the essential
physics of strongly correlated systems.

To conclude, we  have studied the effects of long range coulomb interactions on the
Mott transition using  an extended DMFT. 
 We find that the coulomb interactions  dynamically 
screen the effective on site
interaction.
We find that
to lowest order, the screening is determined by the low energy scale $w$
which is related to the  width of the Kondo
resonance at the Fermi level.  Consequently, the screening is zero in the
insulator. 
 Extending our analysis to three dimensions, we find that the screening
term is non-analytic in $w$ resulting in non-analytic terms in the
free energy.
As a result the Mott transition which was continuous in the absence of these
dynamical screening terms, now becomes a first order transition vindicating
Mott's ideas.

\end{multicols}


\begin{thebibliography}{99}
\bibitem{nrev} M. Imada, A. Fujimori and T. Tokura, {\it Metal Insulator
Transitions}  Rev. Mod. Phys. {\bf 70}, 1039 (1998).
\bibitem{mott}
N.F. Mott, Phil. Mag., {\bf 6} (1961) 287.

\bibitem{hub}
J. Hubbard, Proc. Roy. Soc. (London) A {\bf 281}, 401 (1964).
\bibitem{slater} J.C. Slater, Phys. Rev. {\bf 82},538 (1951).

\bibitem{br}
W.F. Brinkman and T.M. Rice, Phys. Rev. B {\bf 2}, 4302 (1970).

\bibitem{giamarchi}
C. M. Varma and 
T. Giamarchi in {\it Strongly Interacting Fermions and High Tc 
Superconductivity} B. Doucot and J. Zinn Justin editors  Elsevier 1995.

\bibitem{rev}
A.Georges, G. Kotliar, W. Krauth and M. Rozenberg, Rev. Mod. Phys. {\bf 68},
13 (1996).

\bibitem{mott1} N.F. Mott, {\it Metal Insulator Transitions}, Taylor \&
Francis, 1974.

\bibitem{goetz}
G. Moeller, Q. Si, G. Kotliar, M.J. Rozenberg and D.S. Fisher,
Phys. Rev. Lett., {\bf 74}, 2082 (1995).

\bibitem{note}
G. Kotliar, Rutgers University preprint.

\bibitem{kajueter}
H. Kajueter and G. Kotliar, Rutgers University preprint.
\bibitem{Si}
Q. Si and J. L Smith Phys. Rev. Lettt 77, 3391.
J. Lleweilun Smith and Q. Si, cond-mat 9903083.


\bibitem{marcelo}
A. Georges and W. Krauth, Phys. Rev. B {\bf 48}, 7167 (1993);
M.J. Rozenberg, G. Kotliar and X.Y. Zhang, Phys. Rev. B {\bf 49}, 10181 (1994).


\end{thebibliography}
\end{document}